\documentclass[aps,prd,twocolumn,nofootinbib,showpacs]{revtex4}

\usepackage{graphicx}

\usepackage[colorlinks=true, linkcolor=blue, citecolor=blue, urlcolor=blue]{hyperref}

\begin{document}

\title{Reanalysis of the top-quark pair hadroproduction and a precise determination of the top-quark pole mass at the LHC}

\author{Sheng-Quan Wang$^{1}$}
\email[email:]{sqwang@cqu.edu.cn}

\author{Xing-Gang Wu$^{2}$}
\email[email:]{wuxg@cqu.edu.cn}

\author{Jian-Ming Shen$^{3}$}
\email[email:]{shenjm@hnu.edu.cn}

\author{Stanley J. Brodsky$^{4}$}
\email[email:]{sjbth@slac.stanford.edu}

\address{$^{1}$Department of Physics, Guizhou Minzu University, Guiyang 550025, P.R. China}
\address{$^{2}$Department of Physics, Chongqing University, Chongqing 401331, P.R. China}
\address{$^{3}$School of Physics and Electronics, Hunan University, Changsha 410082, P.R. China}
\address{$^{4}$SLAC National Accelerator Laboratory, Stanford University, Stanford, California 94039, USA}

\begin{abstract}
In this paper, we calculate the $t\bar{t}$ pQCD production cross-section at NNLO and determine the top-quark pole mass from recent measurements at the LHC at $\sqrt{S}=13$ TeV center-of-mass energy to high precision by applying the Principle of Maximum Conformality (PMC). The PMC provides a systematic method which rigorously eliminates QCD renormalization scale ambiguities by summing the nonconformal $\beta$ contributions into the QCD coupling constant.  The PMC predictions satisfy the requirements of renormalization group invariance, including renormalization scheme independence, and the PMC scales accurately reflect the virtuality of the underlying production subprocesses.  By using the PMC, an improved prediction for the $t\bar{t}$ production cross-section is obtained without scale ambiguities, which in turn provides a precise value for the top-quark pole mass. Moreover, the predictive power of PMC calculations that the magnitude of higher-order PMC predictions are well within the error bars predicted from the known lower-order has been demonstrated for the top-quark pair production. The resulting determination of the top-quark pole mass $m_t^{\rm pole}=172.5\pm1.4$ GeV from the LHC measurement at $\sqrt{S}=13$ TeV is in agreement with the current world average cited by the Particle Data Group (PDG).  The PMC prediction provides an important high-precision test of the consistency of pQCD and the SM at $\sqrt{S}=13$ TeV with previous LHC measurements at lower CM energies.
\end{abstract}

\pacs{12.38.Aw, 11.10.Gh, 11.15.Bt, 14.65.Ha}

\maketitle

\section{Introduction}
\label{sec:1}

The top-quark was discovered in 1995 by the CDF and D0 Collaborations~\cite{Abe:1995hr, D0:1995jca}.   The large mass of the top-quark and its large Yukawa coupling to the Higgs boson plays a crucial role in testing the electroweak symmetry breaking mechanism and for searching for new physics beyond the SM. Due to its large mass, the top-quark has a short lifetime, decaying well before hadronization takes place.  The spin of the top-quark is transferred directly to its decay products, which provides a unique platform for studying its primary QCD interactions.  The determination of the value of the top-quark mass to high precision is thus of great interest.

The top-quark mass is also a primary input parameter of the SM. For example,
\begin{itemize}
\item
The stability of the quantum vacuum derived  from the shape of the Higgs potential is very sensitive to the top-quark mass;  a precise value of the top-quark mass is thus required in order to accurately predict the evolution of  vacuum stability~\cite{Degrassi:2012ry, Alekhin:2012py}.
\item
The couplings of the top-quark to other particles are fixed through the gauge structure in the SM. The top-quark mass is related to the W-boson mass and the Higgs-boson mass through radiative and loop corrections. A precise determination of these three masses thus gives an important test of the internal consistency of the SM.
\end{itemize}

Currently, the top-quark mass is inferred in two basic ways:  The first approach employs kinematic observables sensitive to the top-quark mass; e.g., one can determine the top-quark mass from the kinematic reconstruction of the top-quark's decay products. Such top-quark mass measurements are referred to as ``MC mass" ($m_t^{\rm MC}$); the most precise determinations of the top-quark mass have been obtained in this approach. Recently, the CMS and ATLAS collaborations at the LHC have established the value $m_t^{\rm MC}=172.26\pm0.61$ GeV~\cite{Sirunyan:2018mlv}. However, these direct kinematical determinations are not linked to the QCD Lagrangian top-quark mass in the specific renormalization scheme employed in theoretical predictions.

An alternative approach employs the mass dependence of the cross-section calculated at next-to-next order (NNLO) perturbative QCD (pQCD).  The top-quark mass can be determined by comparing the measured cross-section with the fixed-order theoretical predictions. This method allows for extractions of the top-quark mass in theoretically well-defined mass schemes, and the extracted mass can then be identified with the top-quark pole mass ($m_t^{\rm pole}$). Theoretical arguments suggest that the difference between the MC mass $m_t^{\rm MC}$ and the pole mass $m_t^{\rm pole}$ is about 1 GeV~\cite{Buckley:2011ms, Hoang:2018zrp, Nason:2016tiy}. It is noted that quarks are confined inside hadrons and have not been observed as physical particles. The notion of a quark mass relies on a theoretical definition. The pole mass is the renormalized quark mass in the on-shell (OS) scheme for the renormalization of the QCD Lagrangian.

Much effort has been devoted to determining the pole mass $m_t^{\rm pole}$ by comparing measurements with the predicted $t\bar{t}$ production cross-section (see e.g., \cite{Aad:2014kva, Aad:2019hzw, Khachatryan:2016mqs, Sirunyan:2017uhy, Sirunyan:2018goh, Abazov:2011pta, Cooper-Sarkar:2020twv, Alekhin:2017kpj}). The Particle Data Group (PDG) currently gives the world average of the top-quark pole mass~\cite{Zyla:2020oca}:
\begin{eqnarray}
m_t^{\rm pole}=172.4\pm0.7~{\rm GeV}.
\end{eqnarray}

In order to provide  maximal constraints on the top-quark pole mass, a key goal is to obtain a highly precise theoretical prediction for the top-quark pair production cross-section. The QCD prediction depends in detail on the choice of the renormalization scale $\mu_r$ controlling the QCD running coupling $\alpha_s(Q^2).$  It has been conventional to either guess the renormalization scale in order to represent the characteristic momentum flow $Q$ of the pQCD process, or to minimize the large logarithmic corrections in the pQCD series.  For example, the renormalization scale can be chosen as the top-quark mass $m_{t}$ in order to eliminate large logarithmic terms such as $\ln(\mu_r/m_{t})$; the  uncertainty from theory is then estimated by varying the guessed renormalization scale over an arbitrary range, e.g., $\mu_r\in[m_{t}/2, 2m_{t}]$.  This uncertainty in determining the renormalization scale is the main source of the uncertainty of the predicted top-quark pair production cross-section and thus the extracted top-quark pole mass.

An essential principle of Renormalization Group Invariance (RGI), is that a physical observable cannot depend on theoretical conventions such as the choice of the renormalization scheme or the renormalization scale.  The conventional procedure of guessing the renormalization scale introduces an inherent scheme-and-scale dependence for the pQCD predictions, it thus violates the fundamental principle of RGI.  Moreover, the perturbative series based on a guessed scale is in general factorially divergent at large orders, behaving as $\alpha_s^n \beta_0^n n!$--the ``renormalon" problem~\cite{Beneke:1998ui} where the $\beta_n$ determine the logarithmic evolution of $\alpha_s$.  Furthermore, the theoretical error estimated by simply varying $\mu_r$ over an arbitrary range is clearly an unreliable and arbitrary estimate,  since it only partly reflects the unknown and factorially divergent perturbative contributions from the nonconformal terms, and it has no sensitivity to the conformal contributions.  The conventional procedure of guessing the renormalization scale is also inconsistent with the well-known Gell-Mann-Low procedure~\cite{GellMann:1954fq}, which determines the renormalization scale rigorously and unambiguously in quantum electrodynamics (QED).

The Principle of Maximum Conformality (PMC)~\cite{Wu:2013ei, Brodsky:2011ta, Brodsky:2011ig, Mojaza:2012mf, Brodsky:2013vpa} provides a systematic way to eliminate the renormalization scheme and renormalization scale uncertainties in non-Abelian pQCD predictions. The PMC is underlying principle for the well-known Brodsky-Lepage-Mackenzie approach~\cite{Brodsky:1982gc}, generalizing the BLM procedure to all orders in $\alpha_s$.  The PMC scales are determined at each order in pQCD by simply absorbing all of the $\beta$ terms in the Renormalization Group Equation (RGE) that govern the behavior of the running coupling. After applying PMC scale-setting, the resulting pQCD series matches the corresponding conformal series
with $\beta=0$.  Since the divergent $n!$ renormalon terms do not appear, the convergence of pQCD series is thus greatly improved.  Since the PMC predictions do not depend on an arbitrary choice of the renormalization scheme, PMC scale-setting satisfies all of the principles of RGI~\cite{Brodsky:2012ms, Wu:2014iba, Wu:2019mky}. The application of the PMC for QCD reduces to Gell-Mann-Low scale setting for QED in the Abelian limit ($N_C \to 0$ at fixed $C_F$)  where the running coupling sums all vacuum polarization insertions. After applying the PMC, there is some residual scale dependence due to the uncalculated higher order perturbative terms; however, unlike conventional renormalization scale-setting, this source of theoretical error is highly suppressed~\cite{Wu:2019mky}.

The PMC has been successfully applied to many high energy processes. We have shown that a comprehensive, self-consistent pQCD explanation of both the top-quark pair production cross-section and the top-quark pair forward-backward asymmetry measured at the Tevatron and LHC can be obtained by applying the PMC~\cite{Brodsky:2012rj, Brodsky:2012sz, Brodsky:2012ik, Wang:2014sua, Wang:2015lna}.  Due to the elimination of the renormalization scale ambiguity, the PMC predictions have much less uncertainties compared to the conventional predictions. We have previously obtained precise values for  the top-quark pole mass: $m^{\rm pole}_t=173.7\pm1.5$ GeV and $174.2\pm1.7$ GeV from measurements at the LHC with $\sqrt{s} = 7$ TeV and $8$ TeV, respectively~\cite{Wang:2017kyd}. Highly precise top-quark pair production cross-section have now been measured at the LHC at $\sqrt{S}=13$ TeV (see e.g., \cite{ATLAS2015-049, Aaboud:2016pbd, ATLAS2019-044, Aad:2019hzw, Khachatryan:2015uqb, Khachatryan:2016kzg, Sirunyan:2017uhy, Sirunyan:2018goh, CMS16-013}). The precision of some experimental measurements is even higher than that of theoretical predictions. It is thus of interest to reanalyze the top-quark pair production cross-section and to provide an improved prediction by using the PMC. The top-quark pole mass can then be determined by a detailed comparison of the production cross-section with the experimental measurement given by the LHC with $\sqrt{S}=13$ TeV. The PMC analysis thus provides an important high-precision test of the consistency of pQCD and the SM at $\sqrt{S}=13$ TeV with the LHC measurements at lower CM energies.

The remaining sections of this paper are organized as follows: In Sec.\ref{sec:2}, we calculate the top-quark pair production cross-section by applying the PMC and compare our PMC predictions with the experimental measurement; we discuss uncalculated higher-order contributions for the top-quark pair production cross-section. The resulting precise determination of the top-quark pole mass from the measured $t\bar{t}$ production cross-section at $\sqrt{S}=13$ TeV is presented in Sec.\ref{sec:3} and compared with determinations at lower CM energies.  Section \ref{sec:4} is reserved for a summary.

\section{Numerical results and discussions for the $t\overline{t}$ production cross-section}
\label{sec:2}

According to QCD factorization for inclusive processes at leading twist, the cross-section for the top-quark pair production $p p \to t \bar t X$ can be expressed as the cross-section for the parton-parton subprocess hard scattering process weighted by the parton distribution functions (PDFs) of the partons participating in the scattering processes;  i.e.,
\begin{eqnarray}
\sigma_{H_1 H_2 \to {t\bar{t} X}} = \sum_{i,j} \int\limits_{4m_{t}^2}^{S}\, ds \,\, {\cal L}_{ij}(s, S, \mu_f) \hat \sigma_{ij}(s,\alpha_s(\mu_r),\mu_r,\mu_f),
\end{eqnarray}
where the parton luminosities ${\cal L}_{ij}$,
\begin{displaymath}
{\cal L}_{ij}(s, S, \mu_f) = {1\over S} \int\limits_s^S {d\hat{s}\over \hat{s}} f_{i/H_1}\left(x_1,\mu_f\right) f_{j/H_2}\left(x_2,\mu_f\right).
\end{displaymath}
The parameters $\mu_f$ and $\mu_r$ are the factorization and renormalization scales,  $S$ denotes the hadronic center-of-mass energy squared, and $s=x_1 x_2 S$ is the subprocess center-of-mass energy squared, where $x_1= {\hat{s} / S}$ and $x_2= {s / \hat{s}}$. The PDFs $f_{i/H_{\alpha}}(x_\alpha,\mu_f)$ ($\alpha=1$ or $2$) are the universal functions that describe the probability of finding a parton of type $i$ with light-front momentum fraction between $x_\alpha$ and $x_{\alpha} +dx_{\alpha}$ in the proton $H_{\alpha}$.

The partonic subprocess cross-section $\hat{\sigma}_{ij}$ can be computed order-by-order as a series expansion in powers of $\alpha_s(\mu_r)$. The QCD radiative corrections, up to next-to-next-to-leading order (NNLO), have been calculated in Refs.\cite{Baernreuther:2012ws, Czakon:2012zr, Czakon:2012pz, Czakon:2013goa, Catani:2019iny}. The pQCD  coefficients at NNLO can also be obtained by using the HATHOR program~\cite{Aliev:2010zk} and the Top++ program~\cite{Czakon:2011xx}.

A detailed PMC analysis for the top-quark pair production cross-section up to NNLO level has been given in Refs.\cite{Brodsky:2012rj, Brodsky:2012sz}; we shall not repeat these formulae here. In this paper, we calculate the top-quark pair production cross-section at the LHC with $\sqrt{S}=13$ TeV following a similar procedure. For brevity, we will use $m_t$ to represent $m_t^{\rm pole}$ in the following. To do the numerical calculation, we initially take the top-quark pole mass as $m_t=173.3$ GeV~\cite{toppole} , and utilize the CT14 parton distribution functions~\cite{Dulat:2015mca}. The running coupling is evaluated in the $\overline{\rm MS}$ scheme from $\alpha_s(M_Z)=0.118$.

After applying the PMC, the PMC scales are determined by absorbing all of the $\beta$ terms in the RGE that govern the behavior of the running coupling. The renormalization scale uncertainty is eliminated. The determination of factorization scale is a separate issue and its dependence exists even for conformal theories. However, we have found that the factorization scale uncertainty can be suppressed for the top-quark pair production after using the PMC to obtain the corresponding conformal series with $\beta=0$~\cite{Wang:2014sua}. The factorization scale dependence can be resolved by matching the perturbative prediction with the nonperturbative bound-state dynamics~\cite{Brodsky:2014yha}. At present, we adopt the usual convention of choosing the factorization scale $\mu_f=m_t$.

If one applies conventional scale setting, the total top-quark pair production cross-section is $\sigma_{t\overline{t}}|_{\rm Conv.}=777.7^{+14.6}_{-30.8}$ pb, where its uncertainty is estimated by varying the scale $\mu_r\in[m_t/2, 2m_t]$. The estimated renormalization scale uncertainty for the total top-quark pair production cross-section is relatively small, due to accidental cancelations among  contributing terms at different orders. The renormalization scale uncertainty is rather large for each perturbative term. Thus, fixing the renormalization scale as $m_t$ appears to give a reasonable prediction for the total top-quark pair production cross-section; however, one cannot identify the QCD correction at each perturbative order. For example, we have found that the scale uncertainty of next-to-leading order (NLO) QCD correction terms for the $(q\overline{q})$-channel, which gives the dominant component of the top-quark pair asymmetry, reaches up to $138\%$ by varying the scale $\mu_r\in[m_t/2, 2m_t]$~\cite{Wang:2015lna}. Simply fixing the renormalization scale at $\mu_r=m_t$ also leads to a small top-quark pair asymmetry, well below the data.

In fact, one usually estimates the magnitude of uncalculated higher-order pQCD corrections to the top-quark pair production cross-section by varying the renormalization scale in the range of $\mu_r\in[m_t/2,2m_t]$. This \textit{ad hoc} method only gives a rough estimate of the uncalculated nonconformal $\beta$ terms at higher orders, however, it has no sensitivity to the conformal contributions at the same order. This is because the conformal contributions have no dependence on the renormalization scale. There are some typical examples to illustrate the unreliability of error estimates using this conventional method, for example, the hadroproduction of the Higgs boson via gluon-gluon fusion at NNLO level~\cite{Anastasiou:2002yz, Wang:2016wgw}; the event shape observables in electron-positron collisions~\cite{GehrmannDeRidder:2007hr}, especially, for two classic event shapes the thrust and the C-parameter~\cite{Wang:2019isi}. The top-quark pair production cross-section at LO level also illustrates the unreliability of the conventional error estimates, this will be shown in the following.

\begin{figure}
\begin{center}
\includegraphics[width=0.50\textwidth]{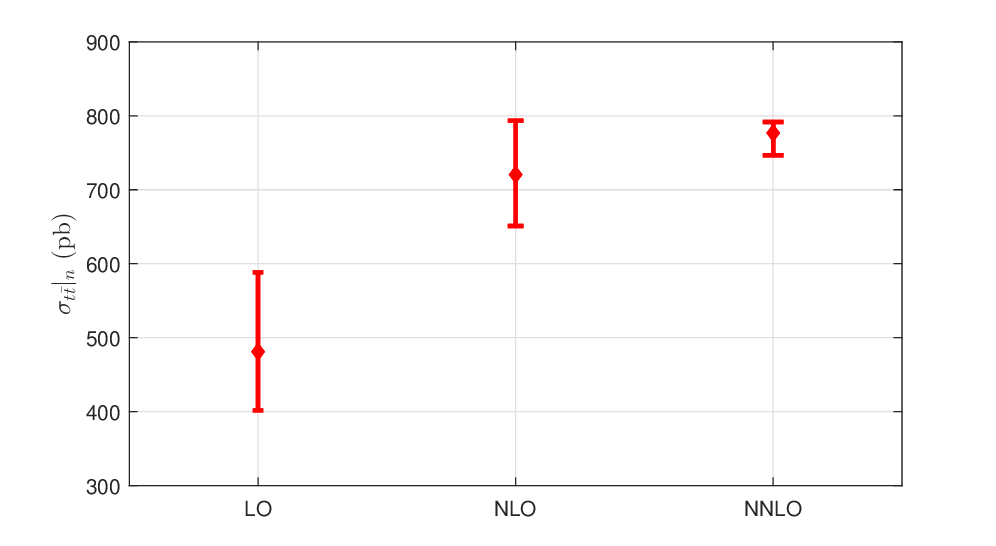}
\caption{The conventional estimates of uncalculated higher-order pQCD corrections, which are obtained by varying $\mu_r\in[m_t/2,2m_t]$ in all known lower-order terms. Here $n=1,2,3$ correspond to the QCD correction at LO, NLO, NNLO level, respectively.}
\label{eshighConv}
\end{center}
\end{figure}

More explicitly, the pQCD corrections for the top-quark pair production cross-section up to $n$th-order level can be expanded in powers of the coupling constant,
\begin{eqnarray}
\sigma_{t\overline{t}}|_n &=& \sum^{n} \limits_{i=\rm {1}} {\cal C}_i\; a_s^{i+1}(\mu_r) \label{Convtt},
\end{eqnarray}
where ${\cal C}_i$ is the $i$th-order QCD correction coefficient, $n=1,2,3$ correspond to the QCD corrections at LO, NLO, NNLO level, respectively. The conventional estimates of uncalculated higher-order pQCD corrections obtained by varying $\mu_r\in[m_t/2,2m_t]$ are presented in Fig.(\ref{eshighConv}). We can see from Figure (\ref{eshighConv}) that the true NNLO cross-section $\sigma_{t\overline{t}}|_{\rm NNLO}$, together with its error bar, is within the predicted NNLO values from the NLO calculation. However, the true NLO cross-section $\sigma_{t\overline{t}}|_{\rm NLO}$, together with its error bar, is outside of the predicted NLO values from the LO calculation. This is also presented by independently varying the renormalization scale and the factorization scale in Ref.\cite{Muselli:2015kba}. Thus, the conventional way of estimating the magnitude of uncalculated higher-order contributions by varying $\mu_r\in[m_t/2,2m_t]$ is invalid for the top-quark pair production at LO level.

After using the PMC scale setting, the situation is different. The PMC scales are determined unambiguously, and they are independent of the choice of the initial scale. The resulting PMC series matches the conformal series with $\beta=0$, a slight change of the PMC scales will break the conformal invariance and thus may lead to large effects~\cite{Chawdhry:2019uuv, Brodsky:2012sz}. We thus cannot simply vary the PMC scales to estimate the magnitude of uncalculated higher-order pQCD corrections, and the conventional way of varying the scales is not applicable to PMC predictions. It is noted an estimate of uncalculated higher-order contributions can be characterized by the convergence of the perturbative QCD series and the magnitude of the last-known higher-order term. We adopt a more conservative method for the estimate of uncalculated higher-order terms for PMC predictions~\cite{Wu:2014iba}. The PMC error estimate is defined to match the value of the contribution from the last-known higher-order term. This error estimate is natural for the PMC scale setting, since after the PMC, the main uncertainty is from the last term at this order with its unfixed PMC scale. Since the renormalization scale uncertainty is rather large for each perturbative term, this approach for the error estimate
cannot be applied to conventional scale setting.

More explicitly, after applying the PMC, the top-quark pair production cross-section (\ref{Convtt}) changes to
\begin{eqnarray}
\sigma_{t\overline{t}}|_n &=& \sum^{n} \limits_{i=\rm {1}} {\tilde{\cal C}}_i\; a_s^{i+1}(Q^{\rm PMC}_i) \label{PMCvtt},
\end{eqnarray}
where ${\tilde{\cal C}}_i$ and $Q^{\rm PMC}_i$ are the $i$th-order conformal coefficient and PMC scale, respectively. The PMC estimate of uncalculated higher-order contributions for an $i$th-order calculation is $\pm|{\tilde{\cal C}}_i\; a_s^{i+1}(Q^{\rm PMC}_i)|_{\rm MAX}$, where both ${\tilde{\cal C}}_i$ and $a_s^{i+1}(Q^{\rm PMC}_i)$ are calculated by varying the scale $\mu_r\in[m_t/2,2m_t]$ and the symbol ``MAX" stands for the maximum value of $|{\tilde{\cal C}}_i\; a_s^{i+1}(Q^{\rm PMC}_i)|$ within this region.

\begin{figure}
\begin{center}
\includegraphics[width=0.50\textwidth]{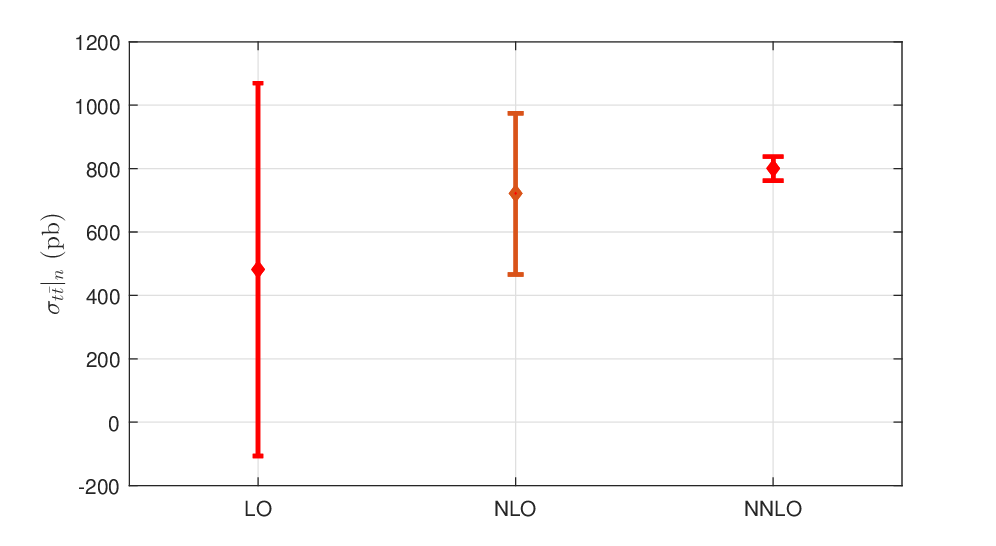}
\caption{The PMC estimates of uncalculated higher-order pQCD corrections, which are taken as $\pm|{\tilde{\cal C}}_i\; a_s^{i+1}(Q^{\rm PMC}_i)|_{\rm MAX}$, where the symbol ``MAX" stands for the maximum value of $|{\tilde{\cal C}}_i\; a_s^{i+1}(Q^{\rm PMC}_i)|$ within the region $\mu_r\in[m_t/2,2m_t]$. Here $n=1,2,3$ correspond to the QCD correction at LO, NLO, NNLO level, respectively.}
\label{eshighPMC}
\end{center}
\end{figure}

The PMC estimates of uncalculated higher-order pQCD corrections for the top-quark pair production cross-section are presented in Fig.(\ref{eshighPMC}). In contrast to the conventional way of varying the scale in the range of $\mu_r\in[m_t/2,2m_t]$, Figure (\ref{eshighPMC}) shows that the higher-order cross sections, together with their error bars, using the PMC are consistently within the predicted cross-sections defined from the lower-order calculations. Moreover, the predicted uncalculated higher-order contributions quickly approach stability. In fact, the predictive power of PMC calculations that the magnitude of higher-order PMC predictions are well within the error bars predicted from the known lower-order has been demonstrated in many high energy processes, for example, the Higgs decay to b-quark pair $\Gamma(\rm {Higgs}\rightarrow b\bar{b})$~\cite{Wu:2014iba} and two gluon $\Gamma(\rm {Higgs}\rightarrow gg)$~\cite{Zeng:2018jzf} processes, the ratio for electron-positron annihilation into hadrons $R(e^{+}e^{-})$~\cite{Wu:2014iba} and the Z boson decay to hadrons $\Gamma(Z\rightarrow \rm{hadron})$ process~\cite{Wang:2014aqa}.

When one applies PMC scale-setting, the renormalization scales at each order are determined by absorbing the $\beta$ terms that govern the behavior of the QCD running coupling via the RGE. The resulting total top-quark pair production cross-section is $\sigma_{t\overline{t}}|_{\rm PMC} = 807.8$ pb for a wide range of the initial choice of scale $\mu_r$. The scale errors for both the total production cross-section and the individual cross-sections at each perturbative order are simultaneously eliminated. Some residual scale dependence will remain due to uncalculated higher-order perturbative terms beyond NNLO.  Unlike the conventional renormalization scale dependence, this scale dependence is negligibly small.

\begin{figure}
\begin{center}
\includegraphics[width=0.50\textwidth]{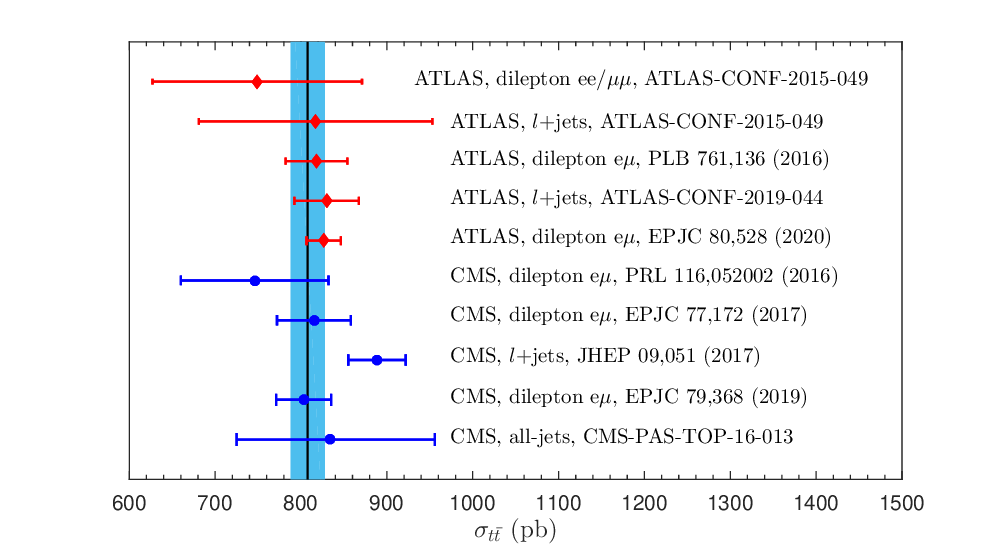}
\caption{A comparison of the PMC prediction with the LHC measurements~\cite{ATLAS2015-049, Aaboud:2016pbd, ATLAS2019-044, Aad:2019hzw, Khachatryan:2015uqb, Khachatryan:2016kzg, Sirunyan:2017uhy, Sirunyan:2018goh, CMS16-013} for the top-quark pair production cross-section at $\sqrt{S}=13$ TeV, where the theoretical error band is estimated by using the CT14 error PDF sets~\cite{Dulat:2015mca} with range of $\alpha_s(M_Z)\in [0.117, 0.119]$ and the uncertainty of $1.5\%$ from the LHC beam energy~\cite{Aaboud:2016pbd}. }
\label{cs13TeVlhc}
\end{center}
\end{figure}

We present a comparison of the PMC prediction with the LHC measurements~\cite{ATLAS2015-049, Aaboud:2016pbd, ATLAS2019-044, Aad:2019hzw, Khachatryan:2015uqb, Khachatryan:2016kzg, Sirunyan:2017uhy, Sirunyan:2018goh, CMS16-013} of the top-quark pair production cross-section at $\sqrt{S}=13$ TeV in Fig.(\ref{cs13TeVlhc}).  The theoretical error band is estimated by using the CT14 error PDF sets~\cite{Dulat:2015mca} with range of $\alpha_s(M_Z)\in [0.117, 0.119]$, as in Ref.\cite{Khachatryan:2016mqs} and the uncertainty of $1.5\%$ from the LHC beam energy~\cite{Aaboud:2016pbd}. Figure (\ref{cs13TeVlhc}) shows that the PMC prediction for the total top-quark pair production cross-section for $\sqrt{S} = 13$ TeV agrees well with all of the corresponding LHC measurements.

\begin{table}[htb]
\centering
\begin{tabular}{|c|c|c|c|c|c|c|}
\hline
~~ ~~  & Measured cross-section ratio\cite{Aad:2019hzw} & PMC prediction \\
\hline
~ $R_{13/7}$ ~ & $4.54\pm0.08\pm0.10\pm0.12$ & $4.61\pm0.15$ \\
\hline
~ $R_{13/8}$ ~& $3.42\pm0.03\pm0.07\pm0.10$ & $3.24\pm0.10$ \\
\hline
\end{tabular}
\caption{The predicted cross-section ratios $R_{13/7}=\sigma^{13\rm TeV}_{t\bar{t}}/\sigma^{7\rm TeV}_{t\bar{t}}$ and $R_{13/8}=\sigma^{13\rm TeV}_{t\bar{t}}/\sigma^{8\rm TeV}_{t\bar{t}}$. The measured cross-section ratios are taken from Ref.~\cite{Aad:2019hzw} as a comparison.
\label{tab1}  }
\end{table}

Since the experimental uncertainties correlated between the two center-of-mass energy cancel out for the ratio of total cross-sections, the more precise results are obtained in comparison to the individual measurements. We calculate the cross-section ratios by the PMC and obtain $R_{13/7}=4.61\pm0.15$ and $R_{13/8}=3.24\pm0.10$, which are presented in Table \ref{tab1}. The predicted cross-section ratios show excellent agreement with the latest ATLAS measurement~\cite{Aad:2019hzw}.

\section{Determination of the top-quark pole mass from the $t\bar{t}$ production cross-section at $\sqrt{S}=13$ TeV}
\label{sec:3}

The top-quark pole mass can be extracted from the comparison of the pQCD prediction of the top-quark pair production cross-section with the corresponding measurements.  Since the precise theoretical prediction for the top-quark pair production cross-section is obtained by using the PMC, we can provide maximal constraints on the top-quark pole mass.

We first parametrize the dependence of the $t\bar{t}$ production cross-section on the top-quark pole mass using the following form~\cite{Beneke:2011mq},
\begin{eqnarray}
\sigma_{t\bar{t}}(m_t)&=&\left(\frac{172.5}{m_t/{\rm GeV}}\right)^4\left(c_0+c_1(\frac{m_t}{\rm GeV}-172.5) \right. \nonumber\\
&&\left.+c_2(\frac{m_t}{\rm GeV}-172.5)^2+c_3(\frac{m_t}{\rm GeV}-172.5)^3\right),
\label{massdepdPMC}
\end{eqnarray}
where the masses are given in units of GeV, and the coefficients $c_{0,1,2,3}$ are determined from the PMC predictions for the top-quark pair cross-section over a wide range of $m_t$~\cite{Wang:2017kyd}.  The renormalization scale uncertainty for the $t\bar{t}$ production cross-sections is eliminated using  the PMC, and thus has less uncertainty compared to the conventional predictions.

In order to extract a reliable top-quark pole mass, we define a likelihood function~\cite{Aaboud:2016000}
\begin{eqnarray}
f(m_t)&=&\int^{+\infty}_{-\infty} f_{\rm th}(\sigma|m_t)\cdot f_{\rm exp}(\sigma|m_t)d\sigma,
\label{likelifunction}
\end{eqnarray}
where the functions $f_{\rm th}(\sigma|m_t)$ and $f_{\rm exp}(\sigma|m_t)$ are normalized Gaussian distributions for the predicted and measured $t\bar{t}$ production cross-sections, respectively. These two functions can be written as
\begin{eqnarray}
f_{\rm th}(\sigma|m_t) = \frac{1}{\sqrt{2\pi}\Delta \sigma_{\rm th}(m_t)}
 \exp \left[-\frac{\left(\sigma-\sigma_{\rm th}(m_t)\right)^2}
 {2\left(\Delta\sigma_{\rm th}(m_t)\right)^2}\right],
\end{eqnarray}
\begin{eqnarray}
f_{\rm exp}(\sigma|m_t) = \frac{1}{\sqrt{2\pi}\Delta \sigma_{\rm exp}(m_t)} \exp{ \left[-\frac{\left(\sigma-\sigma_{\rm exp}(m_t)\right)^2}
 {2\left(\Delta\sigma_{\rm exp}(m_t)\right)^2}\right]}.
\end{eqnarray}
Here, $\sigma_{\rm th}(m_t)$ and $\sigma_{\rm exp}(m_t)$ stand for the predicted and measured $t\bar{t}$ production cross-sections, and the corresponding uncertainties are represented by $\Delta \sigma_{\rm th}(m_t)$ and $\Delta \sigma_{\rm exp}(m_t)$. The central value of the top-quark pole mass is extracted from the maximum of the likelihood function in Eq.(\ref{likelifunction}), and the corresponding error ranges are obtained from the $68\%$ area around the maximum of the likelihood function.

\begin{figure}
\begin{center}
\includegraphics[width=0.50\textwidth]{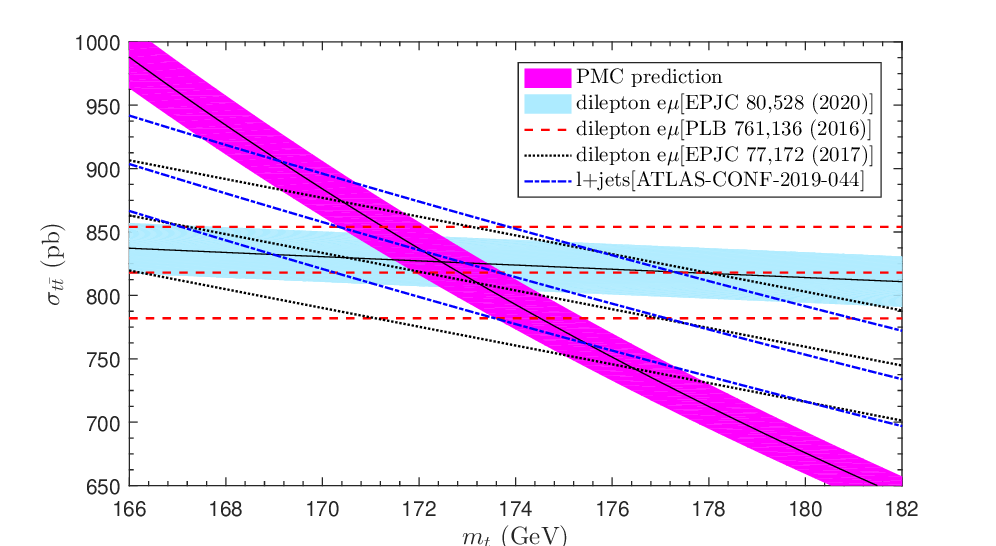}
\caption{The top-quark pair production cross-sections at $\sqrt{S}=13$ TeV as a function of the top-quark pole mass $m_t$, where the thinner shaded band represent the predicted $t\bar{t}$ production cross-section $\sigma_{\rm th}(m_t)$ with the corresponding uncertainty $\Delta \sigma_{\rm th}(m_t)$, which is estimated by using the CT14 error PDF sets with range of $\alpha_s(M_Z)\in [0.117, 0.119]$ and the uncertainty of $1.5\%$ from the LHC beam energy; the thicker shaded band is currently the most precise experimental measurement cross-section $\sigma_{\rm exp}(m_t)$ and the corresponding uncertainty $\Delta \sigma_{\rm exp}(m_t)$~\cite{Aad:2019hzw}. The dashed, dotted, and dash-dot lines are experimental measurements for the dilepton channel~\cite{Aaboud:2016pbd, Khachatryan:2016kzg} and the lepton + jets channel~\cite{ATLAS2019-044}, respectively. The upper and lower lines are the corresponding uncertainties. }
\label{figureTOPde13}
\end{center}
\end{figure}

We present the top-quark pair production cross-sections at $\sqrt{S}=13$ TeV as a function of the top-quark pole mass $m_t$ in Fig.(\ref{figureTOPde13}), where the thinner shaded band represent the predicted $t\bar{t}$ production cross-section $\sigma_{\rm th}(m_t)$ and the corresponding uncertainty $\Delta \sigma_{\rm th}(m_t)$; the thicker shaded band is currently the most precise experimental measurement cross-section $\sigma_{\rm exp}(m_t)$ and the corresponding uncertainty $\Delta \sigma_{\rm exp}(m_t)$~\cite{Aad:2019hzw}. The dashed, dotted, and dash-dot lines are experimental measurements $\sigma_{\rm exp}(m_t)$ for the dilepton channel~\cite{Aaboud:2016pbd, Khachatryan:2016kzg} and the lepton + jets channel~\cite{ATLAS2019-044}, respectively. The upper and lower lines are the corresponding uncertainties. In order to extract a reliable top-quark pole mass, the uncertainty $\Delta \sigma_{\rm th}(m_t)$ should include all the theoretical errors such as $\alpha_s$, PDFs, renormalization and factorization scales. After using the PMC, the renormalization scale uncertainty is eliminated and the factorization scale uncertainty can be suppressed for the top-quark pair production. The $\alpha_s$+PDF uncertainty is estimated by using the CT14 error PDF sets with range of $\alpha_s(M_Z)\in [0.117, 0.119]$, as in Ref.\cite{Khachatryan:2016mqs}. Additionally, the uncertainty of $1.5\%$ from the LHC beam energy is assigned to the predicted cross-section~\cite{Aaboud:2016pbd}.

It is noted that the mass parameter used to characterize the dependence of the measured cross-section on the top-quark mass is the MC mass rather than the pole mass. However, since the mass dependence of the measured cross-sections is very small, as shown by Figure (\ref{figureTOPde13}), and the MC mass and the pole mass differ by about 1 GeV~\cite{Buckley:2011ms, Hoang:2018zrp, Nason:2016tiy}, this approximation causes negligible bias for the determination of the top-quark pole mass. The intersection of the theoretical and experimental curves shown in Figure (\ref{figureTOPde13}) thus gives an unambiguous extraction of the top-quark pole masses.

\begin{table}[htb]
\centering
\begin{tabular}{|c|c|c|c|c|}
\hline
 & dilepton\cite{Khachatryan:2016kzg} & l+jets\cite{ATLAS2019-044} & dilepton\cite{Aaboud:2016pbd} & dilepton\cite{Aad:2019hzw} \\
\hline
 $m_t$~(GeV) & $173.3^{+3.3}_{-3.4}$ & $172.1^{+3.6}_{-3.9}$ & $172.9\pm1.9$ & $172.5\pm1.4$ \\
\hline
\end{tabular}
\caption{Top-quark pole mass (in unit GeV) determined by evaluating the likelihood function (\ref{likelifunction}). The experimental measurements for the dilepton channel~\cite{Aaboud:2016pbd, Khachatryan:2016kzg, Aad:2019hzw} and the lepton + jets channel~\cite{ATLAS2019-044} are taken as the input for $f_{\rm exp}(\sigma|m_t)$.
\label{tab2}  }
\end{table}

By using the experimental measurements for the dilepton channel~\cite{Aaboud:2016pbd, Khachatryan:2016kzg, Aad:2019hzw} and the lepton + jets channel~\cite{ATLAS2019-044} as the input for $f_{\rm exp}(\sigma|m_t)$, and then evaluating the likelihood function (\ref{likelifunction}), the resulting top-quark pole masses are presented in Table \ref{tab2}. It shows that the top-quark pole masses obtained from different channels show good consistency. Due to the measurements for the dilepton channel in Refs.\cite{Aaboud:2016pbd, Aad:2019hzw} are more precise and have less dependence on the top-quark mass compared to the measurements in Refs.\cite{Khachatryan:2016kzg, ATLAS2019-044}, the precision of the top-quark pole masses obtained from the dilepton channel~\cite{Aaboud:2016pbd, Aad:2019hzw} is improved.

\begin{figure}
\begin{center}
\includegraphics[width=0.45\textwidth]{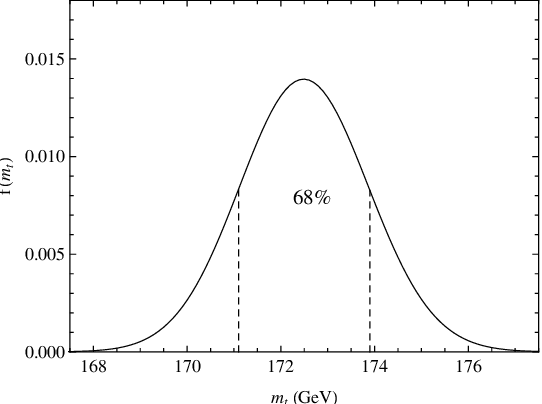}
\caption{The likelihood function $f(m_t)$ defined in Eq.(\ref{likelifunction}) for $\sqrt{S}=$13 TeV, where the area between the two vertical dashed lines stand for the $68\%$ area around the maximum of $f(m_t)$. }
\label{lifunction13TeV}
\end{center}
\end{figure}

For the dilepton channel in Ref.\cite{Aad:2019hzw}, the measured $t\bar{t}$ production cross-section is $\sigma_{t\bar{t}}=(826.4\pm3.6\pm11.5\pm15.7\pm1.9$) pb, with relative uncertainty of $2.4\%$. This cross-section is the most precise result measured so far. By using this measurement as the input for $f_{\rm exp}(\sigma|m_t)$, we present the corresponding likelihood function in Fig.(\ref{lifunction13TeV}), where the area between the two vertical dashed lines stand for the $68\%$ area around the maximum of $f(m_t)$. By evaluating the likelihood function, a reliable top-quark pole mass is extracted to be
\begin{eqnarray}
m_t&=&172.5\pm1.4 ~\rm GeV
\end{eqnarray}
at the LHC for $\sqrt{S}=13$ TeV. The relation between the pole mass and the $\overline{\rm MS}$ mass is currently known up to four-loop level~\cite{Marquard:2015qpa, Kataev:2015gvt}. By converting the top-quark pole mass to the $\overline{\rm MS}$ definition, we obtain
\begin{eqnarray}
m_t^{\overline{\rm MS}}(m_t)&=&162.0\pm1.3 ~\rm GeV
\end{eqnarray}
for $\mu_r=m_t$. By directly applying the PMC to calculate the $\overline{\rm MS}$ mass for the results given in Ref.\cite{Marquard:2015qpa, Kataev:2015gvt}, a precise top-quark $\overline{\rm MS}$ mass is also obtained~\cite{Huang:2020rtx}.

\begin{figure}
\begin{center}
\includegraphics[width=0.50\textwidth]{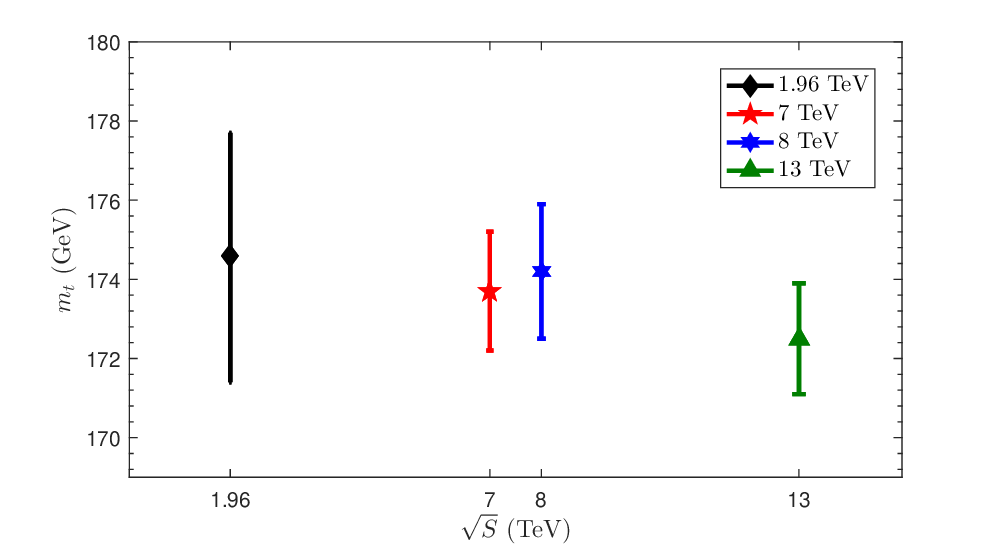}
\caption{The top-quark pole mass determined by the PMC versus the CM energy $\sqrt{S}$, where the PMC results at lower CM energies are take from Ref.\cite{Wang:2017kyd}. }
\label{polemasspmclm}
\end{center}
\end{figure}

Since the experimental uncertainty at the LHC Run II stage with $\sqrt{S}=13$ TeV is smaller than the experimental uncertainty at the LHC Run I stage with $\sqrt{S}=7$ and $8$ TeV, the precision of the determined top-quark pole mass for $\sqrt{S}=13$ TeV is significantly improved compared to the previous analysis for $\sqrt{S}=7$ and $8$ TeV at the LHC and $\sqrt{S}=1.96$ TeV at the Tevatron~\cite{Wang:2017kyd}. We present the top-quark pole mass determined by the PMC versus the CM energy $\sqrt{S}$ in Fig.(\ref{polemasspmclm}), where the PMC results at lower CM energies are take from Ref.\cite{Wang:2017kyd}. A self-consistent determination of the top-quark pole mass can be obtained using the PMC at different CM energies.

The top-quark pole mass is also extracted by comparing the same measured cross-section~\cite{Aad:2019hzw} with the theoretical prediction calculated from conventional scale setting; however, the scale uncertainty is one of the main error sources for the extracted top-quark pole mass. In contrast, since the PMC method eliminates the renormalization scale uncertainty, the determined top-quark pole mass is not plagued by any uncertainty from the choice of the scale $\mu_r$, and thus the precision of the determined top-quark pole mass is improved compared to the result obtained from conventional scale setting~\cite{Aad:2019hzw}.

\begin{figure}
\begin{center}
\includegraphics[width=0.50\textwidth]{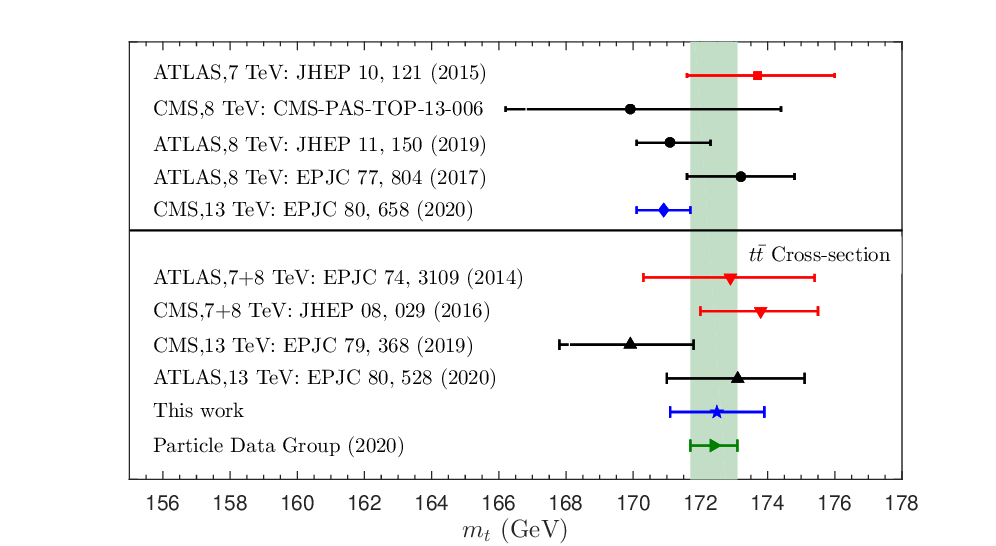}
\caption{A summary of the top-quark pole masses, where our PMC prediction and previous determinations~\cite{Aad:2014kva, Khachatryan:2016mqs, Aad:2019hzw, Sirunyan:2018goh, Aad:2015waa, Aad:2019mkw, CMS13-006, Aaboud:2017ujq, Sirunyan:2019zvx} from collider measurements at different energies and different techniques are presented. The top-quark pole mass, $m_t=172.4\pm0.7$ GeV from the PDG~\cite{Zyla:2020oca} is presented as the shaded band for reference. }
\label{toppolesummary}
\end{center}
\end{figure}

The determined top-quark pole mass using the PMC for $\sqrt{S}=13$ TeV can be cross-checked by other determinations using different techniques by comparing the predicted $t\bar{t}$ cross-sections with the corresponding experimental measurements, including the typical pole masses $m_t=172.9^{+2.5}_{-2.6}$ GeV from ATLAS~\cite{Aad:2014kva} with the $7$ and $8$ TeV data, $m_t=173.8^{+1.7}_{-1.8}$ GeV from CMS~\cite{Khachatryan:2016mqs} with the $7$ and $8$ TeV data, $m_t=173.1^{+2.0}_{-2.1}$ GeV from ATLAS~\cite{Aad:2019hzw} with the $13$ TeV data and $m_t=169.9^{+1.9}_{-2.1}$ GeV from CMS~\cite{Sirunyan:2018goh} with the $13$ TeV data.  In addition to the inclusive $t\bar{t}$ cross-sections, the top-quark pole masses are extracted from the $t\bar{t}$+1jet distribution, yielding $m_t=173.7^{+2.3}_{-2.1}$ GeV from ATLAS~\cite{Aad:2015waa} with the $7$ TeV data, $m_t=171.1^{+1.2}_{-1.0}$ GeV from ATLAS~\cite{Aad:2019mkw} with the $8$ TeV data and $m_t=169.9^{+4.5}_{-3.7}$ GeV from CMS~\cite{CMS13-006} with the $8$ TeV data. The top-quark pole masses are also extracted from the differential distributions, giving $m_t=173.2\pm1.6$ GeV~\cite{Aaboud:2017ujq} and $m_t=170.9\pm0.8$ GeV~\cite{Sirunyan:2019zvx}. More explicitly, we present a summary of the top-quark pole masses in Fig.(\ref{toppolesummary}). The top-quark pole mass, $m_t=172.4\pm0.7$ GeV~\cite{Zyla:2020oca} from the PDG is presented as the shaded band for reference. Figure (\ref{toppolesummary}) shows that the top-quark pole masses obtained from the PMC and the collider measurements at different energies and different techniques show good consistency.

\section{Summary}
\label{sec:4}

Fixed-order pQCD predictions based on conventional scale setting are plagued by the renormalization scale $\mu_r$ uncertainty. In contrast, the PMC provides a rigorous unambiguous method for setting the renormalization scale. The resulting PMC predictions are independent of the choice of the initial renormalization scale and the choice of renormalization scheme. The predictions using the PMC satisfy the principles of RGI.  The PMC is applicable to a wide variety of perturbatively calculable processes. The residual renormalization scale dependence due to uncalculated high-order terms is negligible due to the absence of the renormalon divergence and the convergent pQCD series. The PMC thus greatly improves the precision of tests of the SM.

The conventional way of estimating the magnitude of uncalculated higher-order contributions by varying $\mu_r\in[m_t/2,2m_t]$ is invalid for the top-quark pair production at LO level. After applying the PMC, a comprehensive, self-consistent pQCD explanation for the $t\bar{t}$ production cross-sections measured at the LHC collaborations is obtained. The predictive power of PMC calculations that the magnitude of higher-order PMC predictions are well within the error bars predicted from the known lower-order has been demonstrated for the top-quark pair production. Since the theoretical uncertainty is greatly reduced using the PMC, a reliable determination of the top-quark pole mass  $m_t=172.5\pm1.4$ GeV is obtained by comparing the predicted PMC $t\bar{t}$ cross-section with the latest measurement for $\sqrt{S}=13$ TeV.  Our determination of the top-quark pole mass is consistent with the previous determinations obtained at lower LHC energies and different techniques, giving a new and  important test of the SM. Our determination by applying the PMC is in agreement with the current world average from the PDG, providing complementary information compared to previous determinations.

\hspace{1cm}

{\bf Acknowledgements}: This work was supported in part by the Natural Science Foundation of China under Grants No.11625520, No.11705033, No.11905056 and No.11947406, by the Project of Guizhou Provincial Department under Grant No.KY[2021]003, and by the Department of Energy Contract No.DE-AC02-76SF00515.  SLAC-PUB-17567.


\begin{thebibliography}{99}

\bibitem{Abe:1995hr}
  F.~Abe {\it et al.} [CDF Collaboration],
  Observation of top quark production in $\bar{p}p$ collisions,
  Phys.\ Rev.\ Lett.\  {\bf 74}, 2626 (1995).

\bibitem{D0:1995jca}
  S.~Abachi {\it et al.} [D0 Collaboration],
  Observation of the top quark,
  Phys.\ Rev.\ Lett.\  {\bf 74}, 2632 (1995).

\bibitem{Degrassi:2012ry}
  G.~Degrassi, S.~Di Vita, J.~Elias-Miro, J.~R.~Espinosa, G.~F.~Giudice, G.~Isidori and A.~Strumia,
  Higgs mass and vacuum stability in the Standard Model at NNLO,
  JHEP {\bf 1208}, 098 (2012).

\bibitem{Alekhin:2012py}
  S.~Alekhin, A.~Djouadi and S.~Moch,
  The top quark and Higgs boson masses and the stability of the electroweak vacuum,
  Phys.\ Lett.\ B {\bf 716}, 214 (2012).

\bibitem{Sirunyan:2018mlv}
  A.~M.~Sirunyan {\it et al.} [CMS Collaboration],
  Measurement of the top quark mass in the all-jets final state at $\sqrt{s} =$ 13 TeV and combination with the lepton+jets channel,
  Eur.\ Phys.\ J.\ C {\bf 79}, 313 (2019).

\bibitem{Buckley:2011ms}
  A.~Buckley {\it et al.},
  General-purpose event generators for LHC physics,
  Phys.\ Rept.\  {\bf 504}, 145 (2011).

\bibitem{Hoang:2018zrp}
  A.~H.~Hoang, S.~Pl\"atzer and D.~Samitz,
  On the Cutoff Dependence of the Quark Mass Parameter in Angular Ordered Parton Showers,
  JHEP {\bf 1810}, 200 (2018).

\bibitem{Nason:2016tiy}
  P.~Nason,
  Theory Summary,
  PoS TOP {\bf 2015}, 056 (2016).

\bibitem{Aad:2014kva}
  G.~Aad {\it et al.} [ATLAS Collaboration],
  Measurement of the $t\bar{t}$ production cross-section using $e\mu $ events with b-tagged jets in pp collisions at $\sqrt{s}$ = 7 and 8  $\,\mathrm{TeV}$ with the ATLAS detector,
  Eur.\ Phys.\ J.\ C {\bf 74}, 3109 (2014)
  Addendum: [Eur.\ Phys.\ J.\ C {\bf 76}, 642 (2016)].

\bibitem{Aad:2019hzw}
  G.~Aad {\it et al.} [ATLAS Collaboration],
  Measurement of the $t\bar{t}$ production cross-section and lepton differential distributions in $e\mu $ dilepton events from $pp$ collisions at $\sqrt{s}=13\,\text {TeV}$ with the ATLAS detector,
  Eur.\ Phys.\ J.\ C {\bf 80}, 528 (2020).

\bibitem{Khachatryan:2016mqs}
  V.~Khachatryan {\it et al.} [CMS Collaboration],
  Measurement of the t-tbar production cross section in the e-mu channel in proton-proton collisions at sqrt(s) = 7 and 8 TeV,
  JHEP {\bf 1608}, 029 (2016).

\bibitem{Sirunyan:2017uhy}
  A.~M.~Sirunyan {\it et al.} [CMS Collaboration],
  Measurement of the $t \bar t$ production cross section using events with one lepton and at least one jet in pp collisions at $\sqrt{s}$  = 13 TeV,
  JHEP {\bf 1709}, 051 (2017).

\bibitem{Sirunyan:2018goh}
  A.~M.~Sirunyan {\it et al.} [CMS Collaboration],
  Measurement of the $\mathrm{t}\overline{\mathrm{t}}$ production cross section, the top quark mass, and the strong coupling constant using dilepton events in pp collisions at $\sqrt{s} =$ 13 TeV,
  Eur.\ Phys.\ J.\ C {\bf 79}, 368 (2019).

\bibitem{Abazov:2011pta}
  V.~M.~Abazov {\it et al.} [D0 Collaboration],
  Determination of the pole and $\overline{MS}$ masses of the top quark from the $t\bar{t}$ cross section,
  Phys.\ Lett.\ B {\bf 703}, 422 (2011).

\bibitem{Cooper-Sarkar:2020twv}
  A.~M.~Cooper-Sarkar, M.~Czakon, M.~A.~Lim, A.~Mitov and A.~S.~Papanastasiou,
  Simultaneous extraction of $\alpha_s$ and $m_t$ from LHC $t\bar{t}$ differential distributions,
  arXiv:2010.04171 [hep-ph].

\bibitem{Alekhin:2017kpj}
  S.~Alekhin, J.~Bl$\ddot{\rm u}$mlein, S.~Moch and R.~Placakyte,
  Parton distribution functions, $\alpha_s$, and heavy-quark masses for LHC Run II,
  Phys.\ Rev.\ D {\bf 96}, 014011 (2017).

\bibitem{Zyla:2020oca}
  P.A. Zyla {\it et al.} [Particle Data Group],
  Review of Particle Physics,
  Prog. Theor. Exp. Phys. 2020, 083C01 (2020).

\bibitem{GellMann:1954fq}
  M.~Gell-Mann and F.~E.~Low,
  Quantum electrodynamics at small distances,
  Phys.\ Rev.\  {\bf 95}, 1300 (1954).

\bibitem{Beneke:1998ui}
  M.~Beneke,
  Renormalons,
  Phys.\ Rept.\  {\bf 317}, 1 (1999).

\bibitem{Wu:2013ei}
  X.~G.~Wu, S.~J.~Brodsky and M.~Mojaza,
  The Renormalization Scale-Setting Problem in QCD,
  Prog.\ Part.\ Nucl.\ Phys.\  {\bf 72}, 44 (2013).

\bibitem{Brodsky:2011ta}
  S.~J.~Brodsky and X.~G.~Wu,
  Scale Setting Using the Extended Renormalization Group and the Principle of Maximum Conformality: the QCD Coupling Constant at Four Loops,
  Phys.\ Rev.\ D {\bf 85}, 034038 (2012)
  [Phys.\ Rev.\ D {\bf 86}, 079903 (2012)].

\bibitem{Brodsky:2011ig}
  S.~J.~Brodsky and L.~Di Giustino,
  Setting the Renormalization Scale in QCD: The Principle of Maximum Conformality,
  Phys.\ Rev.\ D {\bf 86}, 085026 (2012).

\bibitem{Mojaza:2012mf}
  M.~Mojaza, S.~J.~Brodsky and X.~G.~Wu,
  Systematic All-Orders Method to Eliminate Renormalization-Scale and Scheme Ambiguities in Perturbative QCD,
  Phys.\ Rev.\ Lett.\  {\bf 110}, 192001 (2013).

\bibitem{Brodsky:2013vpa}
  S.~J.~Brodsky, M.~Mojaza and X.~G.~Wu,
  Systematic Scale-Setting to All Orders: The Principle of Maximum Conformality and Commensurate Scale Relations,
  Phys.\ Rev.\ D {\bf 89}, 014027 (2014).

\bibitem{Brodsky:1982gc}
 S.~J.~Brodsky, G.~P.~Lepage and P.~B.~Mackenzie,
 On the Elimination of Scale Ambiguities in Perturbative Quantum Chromodynamics,
 Phys.\ Rev.\ D {\bf 28}, 228 (1983).

\bibitem{Brodsky:2012ms}
  S.~J.~Brodsky and X.~G.~Wu,
  Self-Consistency Requirements of the Renormalization Group for Setting the Renormalization Scale,
  Phys.\ Rev.\ D {\bf 86}, 054018 (2012).

\bibitem{Wu:2014iba}
  X.~G.~Wu, Y.~Ma, S.~Q.~Wang, H.~B.~Fu, H.~H.~Ma, S.~J.~Brodsky and M.~Mojaza,
  Renormalization Group Invariance and Optimal QCD Renormalization Scale-Setting,
  Rept.\ Prog.\ Phys.\  {\bf 78}, 126201 (2015).

\bibitem{Wu:2019mky}
  X.~G.~Wu, J.~M.~Shen, B.~L.~Du, X.~D.~Huang, S.~Q.~Wang and S.~J.~Brodsky,
  The QCD Renormalization Group Equation and the Elimination of Fixed-Order Scheme-and-Scale Ambiguities Using the Principle of Maximum Conformality,
  Prog.\ Part.\ Nucl.\ Phys.\  {\bf 108}, 103706 (2019).

\bibitem{Brodsky:2012rj}
  S.~J.~Brodsky and X.~G.~Wu,
  Eliminating the Renormalization Scale Ambiguity for Top-Pair Production Using the Principle of Maximum Conformality,
  Phys.\ Rev.\ Lett.\  {\bf 109}, 042002 (2012).

\bibitem{Brodsky:2012sz}
  S.~J.~Brodsky and X.~G.~Wu,
  Application of the Principle of Maximum Conformality to Top-Pair Production,
  Phys.\ Rev.\ D {\bf 86}, 014021 (2012)
  [Phys.\ Rev.\ D {\bf 87}, 099902 (2013)].

\bibitem{Brodsky:2012ik}
  S.~J.~Brodsky and X.~G.~Wu,
  Application of the Principle of Maximum Conformality to the Top-Quark Forward-Backward Asymmetry at the Tevatron,
  Phys.\ Rev.\ D {\bf 85}, 114040 (2012).

\bibitem{Wang:2014sua}
  S.~Q.~Wang, X.~G.~Wu, Z.~G.~Si and S.~J.~Brodsky,
  Application of the Principle of Maximum Conformality to the Top-Quark Charge Asymmetry at the LHC,
  Phys.\ Rev.\ D {\bf 90}, 114034 (2014).

\bibitem{Wang:2015lna}
  S.~Q.~Wang, X.~G.~Wu, Z.~G.~Si and S.~J.~Brodsky,
  Predictions for the Top-Quark Forward-Backward Asymmetry at High Invariant Pair Mass Using the Principle of Maximum Conformality,
  Phys.\ Rev.\ D {\bf 93}, 014004 (2016).

\bibitem{Wang:2017kyd}
  S.~Q.~Wang, X.~G.~Wu, Z.~G.~Si and S.~J.~Brodsky,
  A precise determination of the top-quark pole mass,
  Eur.\ Phys.\ J.\ C {\bf 78}, 237 (2018).

\bibitem{ATLAS2015-049}
  ATLAS Collaboration,
  Measurements of the $t\bar{t}$ production cross-section in the dilepton and lepton-plus-jets channels and of the ratio of the $t\bar{t}$ and $Z$ boson cross-sections in $pp$ collisions at $\sqrt{s}=13$ TeV with the ATLAS detector,
  ATLAS-CONF-2015-049.

\bibitem{Aaboud:2016pbd}
  M.~Aaboud {\it et al.} [ATLAS Collaboration],
  Measurement of the $t\bar{t}$ production cross-section using $e\mu$ events with b-tagged jets in pp collisions at $\sqrt{s}$=13 TeV with the ATLAS detector,
  Phys.\ Lett.\ B {\bf 761}, 136 (2016)
  Erratum: [Phys.\ Lett.\ B {\bf 772}, 879 (2017)].

\bibitem{ATLAS2019-044}
  ATLAS Collaboration,
  Measurement of the $t\bar{t}$ production cross-section in the lepton+jets channel at $\sqrt{s}=13$ TeV with the ATLAS experiment,
  ATLAS-CONF-2019-044.

\bibitem{Khachatryan:2015uqb}
  V.~Khachatryan {\it et al.} [CMS Collaboration],
  Measurement of the top quark pair production cross section in proton-proton collisions at $\sqrt(s) =$ 13 TeV,
  Phys.\ Rev.\ Lett.\  {\bf 116}, 052002 (2016).

\bibitem{Khachatryan:2016kzg}
  V.~Khachatryan {\it et al.} [CMS Collaboration],
  Measurement of the $t\bar{t}$ production cross section using events in the e$\mu$ final state in pp collisions at $\sqrt{s} =$ 13 TeV,
  Eur.\ Phys.\ J.\ C {\bf 77}, 172 (2017).

\bibitem{CMS16-013}
  CMS Collaboration,
  Measurement of the $t\bar{t}$ production cross section at 13 TeV in the all-jets final state,
  CMS-PAS-TOP-16-013.

\bibitem{Baernreuther:2012ws}
  P.~Baernreuther, M.~Czakon and A.~Mitov,
  Percent Level Precision Physics at the Tevatron: First Genuine NNLO QCD Corrections to $q \bar{q} \to t \bar{t} + X$,
  Phys.\ Rev.\ Lett.\  {\bf 109}, 132001 (2012).

\bibitem{Czakon:2012zr}
  M.~Czakon and A.~Mitov,
  NNLO corrections to top-pair production at hadron colliders: the all-fermionic scattering channels,
  JHEP {\bf 1212}, 054 (2012).

\bibitem{Czakon:2012pz}
  M.~Czakon and A.~Mitov,
  NNLO corrections to top pair production at hadron colliders: the quark-gluon reaction,
  JHEP {\bf 1301}, 080 (2013).

\bibitem{Czakon:2013goa}
  M.~Czakon, P.~Fiedler and A.~Mitov,
  Total Top-Quark Pair-Production Cross Section at Hadron Colliders Through $O(\alpha^4_S)$,
  Phys.\ Rev.\ Lett.\  {\bf 110}, 252004 (2013).

\bibitem{Catani:2019iny}
  S.~Catani, S.~Devoto, M.~Grazzini, S.~Kallweit, J.~Mazzitelli and H.~Sargsyan,
  Top-quark pair hadroproduction at next-to-next-to-leading order in QCD,
  Phys.\ Rev.\ D {\bf 99}, 051501 (2019).

\bibitem{Aliev:2010zk}
  M.~Aliev, H.~Lacker, U.~Langenfeld, S.~Moch, P.~Uwer and M.~Wiedermann,
  HATHOR: HAdronic Top and Heavy quarks crOss section calculatoR,
  Comput.\ Phys.\ Commun.\  {\bf 182}, 1034 (2011).

\bibitem{Czakon:2011xx}
  M.~Czakon and A.~Mitov,
  Top++: A Program for the Calculation of the Top-Pair Cross-Section at Hadron Colliders,
  Comput.\ Phys.\ Commun.\  {\bf 185}, 2930 (2014).

\bibitem{toppole}
  The ATLAS and CMS Collaborations,
  Combination of ATLAS and CMS results on the mass of the top quark using up to 4.9 fb$^{-1}$ of data,
  ATLAS-CONF-2012-095, CMS-PAS-TOP-12-001.

\bibitem{Dulat:2015mca}
  S.~Dulat, {\it et al.},
  New parton distribution functions from a global analysis of quantum chromodynamics,
  Phys.\ Rev.\ D {\bf 93}, 033006 (2016).

\bibitem{Brodsky:2014yha}
  S.~J.~Brodsky, G.~F.~de Teramond, H.~G.~Dosch and J.~Erlich,
  Light-Front Holographic QCD and Emerging Confinement,
  Phys.\ Rept.\  {\bf 584}, 1 (2015).

\bibitem{Anastasiou:2002yz}
  C.~Anastasiou and K.~Melnikov,
  Higgs boson production at hadron colliders in NNLO QCD,
  Nucl.\ Phys.\ B {\bf 646}, 220 (2002).

\bibitem{Wang:2016wgw}
  S.~Q.~Wang, X.~G.~Wu, S.~J.~Brodsky and M.~Mojaza,
  Application of the Principle of Maximum Conformality to the Hadroproduction of the Higgs Boson at the LHC,
  Phys.\ Rev.\ D {\bf 94}, 053003 (2016).

\bibitem{GehrmannDeRidder:2007hr}
  A.~Gehrmann-De Ridder, T.~Gehrmann, E.~W.~N.~Glover and G.~Heinrich,
  NNLO corrections to event shapes in e+ e- annihilation,
  JHEP {\bf 0712}, 094 (2007).

\bibitem{Wang:2019isi}
  S.~Q.~Wang, S.~J.~Brodsky, X.~G.~Wu, J.~M.~Shen and L.~Di Giustino,
  Novel method for the precise determination of the QCD running coupling from event shape distributions in electron-positron annihilation,
  Phys.\ Rev.\ D {\bf 100}, 094010 (2019).

\bibitem{Muselli:2015kba}
  C.~Muselli, M.~Bonvini, S.~Forte, S.~Marzani and G.~Ridolfi,
  Top Quark Pair Production beyond NNLO,
  JHEP {\bf 1508}, 076 (2015).

\bibitem{Chawdhry:2019uuv}
  H.~A.~Chawdhry and A.~Mitov,
  Ambiguities of the principle of maximum conformality procedure for hadron collider processes,
  Phys.\ Rev.\ D {\bf 100}, 074013 (2019).

\bibitem{Zeng:2018jzf}
  J.~Zeng, X.~G.~Wu, S.~Bu, J.~M.~Shen and S.~Q.~Wang,
  Reanalysis of the Higgs-boson decay $H \to gg$ up to $\alpha_s^6$-order level using the principle of maximum conformality,
  J.\ Phys.\ G {\bf 45}, 085004 (2018).

\bibitem{Wang:2014aqa}
  S.~Q.~Wang, X.~G.~Wu and S.~J.~Brodsky,
  Reanalysis of the Higher Order Perturbative QCD corrections to Hadronic $Z$ Decays using the Principle of Maximum Conformality,
  Phys.\ Rev.\ D {\bf 90}, 037503 (2014).

\bibitem{Beneke:2011mq}
  M.~Beneke, P.~Falgari, S.~Klein and C.~Schwinn,
  Hadronic top-quark pair production with NNLL threshold resummation,
  Nucl.\ Phys.\ B {\bf 855}, 695 (2012).

\bibitem{Aaboud:2016000}
  M.~Aaboud {\it et al.} [ATLAS Collaboration],
  Determination of the Top-Quark Mass from the $t\bar{t}$ Cross Section Measurement in pp Collisions at $\sqrt{S}=7$ TeV with the ATLAS detector,
  ATLAS-CONF-2011-054.

\bibitem{Marquard:2015qpa}
  P.~Marquard, A.~V.~Smirnov, V.~A.~Smirnov and M.~Steinhauser,
  Quark Mass Relations to Four-Loop Order in Perturbative QCD,
  Phys.\ Rev.\ Lett.\  {\bf 114}, 142002 (2015).

\bibitem{Kataev:2015gvt}
  A.~L.~Kataev and V.~S.~Molokoedov,
  On the flavour dependence of the $\mathcal{O}(\alpha_s^4)$ correction to the relation between running and pole heavy quark masses,
  Eur.\ Phys.\ J.\ Plus {\bf 131}, 271 (2016).

\bibitem{Huang:2020rtx}
  X.~D.~Huang, X.~G.~Wu, J.~Zeng, Q.~Yu, X.~C.~Zheng and S.~Xu,
  Determination of the top-quark $\overline{MS}$ running mass via its perturbative relation to the on-shell mass with the help of the principle of maximum conformality,
  Phys.\ Rev.\ D {\bf 101}, 114024 (2020).

\bibitem{Aad:2015waa}
  G.~Aad {\it et al.} [ATLAS Collaboration],
  Determination of the top-quark pole mass using $ t\overline{t} $ + 1-jet events collected with the ATLAS experiment in 7 TeV pp collisions,
  JHEP {\bf 1510}, 121 (2015).

\bibitem{Aad:2019mkw}
  G.~Aad {\it et al.} [ATLAS Collaboration],
  Measurement of the top-quark mass in $t\bar{t}+1$-jet events collected with the ATLAS detector in $pp$ collisions at $\sqrt{s}=8$ TeV,
  JHEP {\bf 1911}, 150 (2019).

\bibitem{CMS13-006}
  CMS Collaboration,
  Determination of the normalised invariant mass distribution of $t\bar{t}$+jet and extraction of the top quark mass,
  CMS-PAS-TOP-13-006.

\bibitem{Aaboud:2017ujq}
  M.~Aaboud {\it et al.} [ATLAS Collaboration],
  Measurement of lepton differential distributions and the top quark mass in $t\bar{t}$ production in $pp$ collisions at $\sqrt{s}=8$ TeV with the ATLAS detector,
  Eur.\ Phys.\ J.\ C {\bf 77}, 804 (2017).

\bibitem{Sirunyan:2019zvx}
  A.~M.~Sirunyan {\it et al.} [CMS Collaboration],
  \mbox{Measurement} of $t\bar t$ normalised multi-differential cross sections in $pp$ collisions at $\sqrt s=13$ TeV, and simultaneous determination of the strong coupling strength, top quark pole mass, and parton distribution functions,
  Eur.\ Phys.\ J.\ C {\bf 80}, 658 (2020).

\end{thebibliography}
\end{document}